# The Boundary Conditions Geometry in Lattice-Ising model

You-gang Feng

Department of Basic Science, College of Science, Gui Zhou University. Cai Jia Guan, Guiyang 550003, China
E-mail: ygfeng45@yahoo.com.cn

We found that the differential topology of the lattice-system of Ising model determines whether there can be the continuous phase transition, the geometric topology of the space the lattice-system is embedded in determines whether the system can become ordered. If the system becomes ordered it may not admit the continuous phase transition. The spin-projection orientations are strongly influenced by the geometric topology of the space the lattice-system is embedded in.

## 1. Introduction

Physicists worked on Ising models for many years have obtained brilliant achievements in the theory of continuous phase transition. The model's charm is that the studying and researching will never be at an end; some new properties and problems will be discovered and explored. Kadanoff pointed out, "The physical problem of critical phenomena reduces to the mathematical problem of enumerating and describing the different universality class."[1] The late Chinese mathematician Xing-sheng Chen, well-known for his great achievements in differential topology around the world, said, "Physics just is geometry, and they are in one family." In this paper we will discuss how the difference between differential and geometric topology influences on the magnetic order of the lattice system of Ising model.

## 2. Applications

It used to be that the discussion about the continuous phase transition of Ising model is concentrated on the property of derivatives of order 2 of the system's free energy at the critical temperature. If it is divergent the system will have magnetic order below the critical temperature, if it is not divergent the ordered state of system will not appear. If we regard the thermodynamic properties of the system as a kind of manifold, and the system's free energy is a function defined in the coordinate system.

The property of derivatives of order 2 of the free energy reflects the differential topology of the manifold. The phase transitions of the systems with the same differential topology belong to one kind of phase transitions—a phase transition of the second kind. The order parameters are actually the invariance linked to the differential topology of such systems. Consequently, the discussion about the phase transition of the Ising model has a meaning in terms of differential topology, although people did not pay much attention to relate the phase-transition property to the property of differential topology, and were willing to consider it as a physical property only. As a result of this, they were so indifferent to the geometric topological property of the space in which the lattice system is embedded, that the selection of spin-projection orientations has never been mentioned. This aspect is dependent upon the geometric topology of the space. Considering the spin-projection orientations, the lattice gas will not be equivalent to the Ising model because the lattice gas itself has not any projecting orientation [2].

The property of differential topology and the property of geometry topology are two distinct properties of topology: the former determines whether there can exist the continuous phase transition in a lattice system, and the later determines whether the system can become ordered. Being a continuous phase transition, the system will certainly become ordered, which was proved by physicists. But if the system behaves ordered, it may not be in the state of continuous phase transition for some systems. This results from the difference between the two topologies above mentioned. So it is necessary to discuss the influence of the geometric topology on the magnetic order of the Ising model. We will discuss the problems in the following three respects.

First, the geometric topology of the embedding space for the lattice system will restrict the spin-projection orientations. A key point of the Onsager's solution is his periodic boundary conditions. Under the conditions, as illustrated in FIG .1, the two

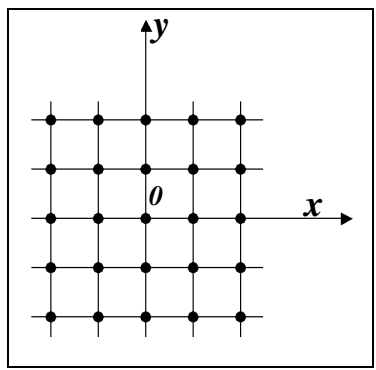

FIG. 1

lattices in sites (-∞, y) or (x,-∞) and (+∞, y) or (x,+∞) have the same spin state.

The periodic boundary conditions change the plane square-lattice system into a torus-lattice system, which should be embedded in 3-dimensional Euclidean space. If a lattice system can become ordered, it will be in one of the following two states: The entire lattice system can be referred to as a lattice with total spin. So, the space is simply connected; or there is a non-vanishing vector field, of which the appearance is dependent upon the geometric topology of the lattice system. The torus-lattice system is not contractible because of its geometric topological structure[3]. FIG .2 shows that three simple closed curves on the torus, and only one's inner block of which, the curve B inner block, can become contractible. The singularity of the Onsager's equation indicated that the system certainly has the continuous phase transition, which implied the system is ordered but contractible. We noticed that there are only two spin states for the Ising model: spin-up and spin-down, and there are only three important spin-projection orientations in various projecting orientations: the directions normal to the plane, parallel to the x-axis and to the y-axis as illustrated in FIG.1. In fact, the partition function of the system is not affected by the projecting orientations because there is no the spin-projection orientation term in it. We found that the spin-projection orientation normal to the plane with the periodic boundary conditions is forbidden. The reason is that if the system were ordered, the total spin-projection orientation of the system should be normal to the torus surface, which cannot shrink. Then its projecting orientation is uncertain because the normal orientations of the torus are different and divergent everywhere.

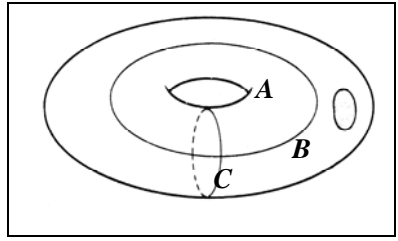

FIG.2

However, if the spin-projection orientations are parallel to the x-axis or to the y-axis, the situation is completely different. Let the total spin of the system be $S$ below its critical temperature, the average lattice spin be $s$, and $s = S/N$, $N$ be the total number of the lattices in the system, $N \to \infty$, and $s \neq 0$, in accord with the thermodynamic limit condition. When the torus-lattice system becomes ordered, there is a continuous non-vanishing vector field on the torus, as illustrated in FIG 3 each little arrow represents an average lattice spin, which means that the topological property of torus allows the system to be ordered. It is the geometric topology that not only determines system is ordered, but also restricts the spin-projection orientations. Onsager's solution means that the torus system's differential topology is equivalent to the plane system's differential topology, but the principle of topology tells us that their geometry topology are different. Notwithstanding they cannot change into each other automatically, (i.e., they are not one-to-one).

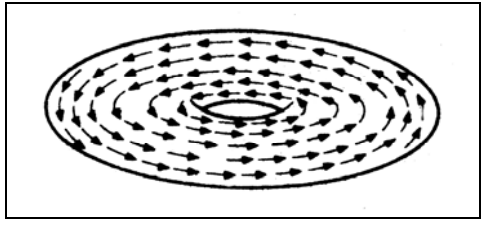

FIG . 3

In fact, the free energy of the torus-lattice system is different from the free- energy of the plane square-lattice system. The partition function of the latter is defined as

$$Q = \sum_{s_i} \exp(-\frac{H}{k_B T}) = \sum_{s_i}^{(1)} \exp(-\frac{H}{k_B T}) + \sum_{s_i}^{(2)} \exp(-\frac{H}{k_B T}) \qquad (1)$$

$$H = -J \sum_{i,j} s_i s_j, \qquad (2)$$

where $H$ is the system Hamiltonian, $J$ is a spin-coupling constant, $\sum_{i,j}$ denotes the sum over all possible nearest neighbor lattices, $\sum_{s_i}$ does the sum over all possible states of spins, $\sum_{s_i}^{(1)}$ is the sum over all possible states of spins with the periodic boundary conditions, and $\sum_{s_i}^{(2)}$ denotes the sum over all possible states of spins which fail to keep the conditions. The first term on the right hand of Eq.(1) is expressed by

$$Q_1 = \sum_{s_i}^{1} \exp(-\frac{H}{k_B T}) \qquad (3)$$

Obviously, $Q_1$ is the partition function of the lattice system with the periodic boundary conditions, namely, the partition function of the torus-lattice system. The free-energy of the plane square-lattice system and the free-energy of the torus-lattice system are respectively given by

$$F = -k_B T \ln Q \tag{4}$$

$$F_1 = -k_B T \ln Q_1 \tag{5}$$

Because of the exponential function property, we have

$$Q > 0, \qquad Q_1 > 0, \qquad Q > Q_1 \tag{6}$$

the difference between $F_1$ and $F$ is

$$\Delta F = F_1 - F = k_B T \ln(\frac{Q}{Q_1}) > 0 \tag{7}$$

Under the thermodynamic limit condition $\Delta F$ changes into infinitesimal but non-zero, which makes the critical temperature of the torus-lattice system infinitely close to the critical temperature of the plane square-lattice system so that the Onsager's solution can be regarded as an exact solution of the plane square-lattice system. The infinitesimal difference $\Delta F > 0$ shows that the original state cannot change into the state with the periodic boundary conditions in the thermodynamic equilibrium state, and the small offset of the free-energy makes the transformation of the topological structure of the system so great that its fundamental group is completely different from the original[3]. The thermodynamic limit guarantees the systems with and without the periodic boundary conditions have the same property of differential topology so as to have the same property of the continuous phase transition. This is not only seen in the plane-lattice system, but also will be seen in the one-dimensional lattice system [2]. Furthermore, if the spin states of all lattices at the boundary are the same (which is one kind of the periodic boundary conditions) these lattices themselves will gather up to one lattice to make the plane-lattice system become a sphere-lattice system. The solution of such a system will also be equivalent to Onsager's solution because of its periodic boundary conditions, and there exists certainly the continuous phase transition. Its geometric topology shows the sphere surface is simply connected[3], which allows the system to be ordered. However, there is no non-vanishing vector field on the sphere surface because of the "hairy ball theorem"[3], which is distinct from the torus-lattice system. When the sphere-lattice system becomes ordered its total spin-projection orientation will be along the sphere radial, which is arbitrary.

The second sense of the influence of the geometric properties is that the ordered system is not equivalent to its appearance of continuous phase transition all the time. A one-dimensional Ising model can be constructed on each site, which coordinate positions are $x = 0,+1,-1,+2,-2,....,$ on which is laid one lattice spin with free degree $n = 1$. Such a system can be referred as to a Ising model on the real line. Using the periodic boundary conditions, it is proved that the 1-dimensional system has not the continuous phase transition. The periodic boundary conditions changed the real line into a circle embedded in a plane, which means that the circle-lattice system also cannot have the continuous phase transition like the real-line-lattice system because of the the thermodynamic limit condition. However, the real-line–lattice system can become ordered because of its simply connected[3]. In other words, the whole lattice system can be looked as a lattice with the total spin, but there is no non-vanishing vector field. The circle is not simply connected, and it allows the non-vanishing vector field to exist, which direction is clockwise or anticlockwise. Here, we noticed that the 1-dimensional lattice system with or without the periodic boundary conditions have the same property of differential topology but the geometric topology makes them appear in different spin-projection orientations. The key point to cause the system to become ordered is its geometric topology, not its differential topology. While the free degree of spin changes from $n = 1$ into $1 \langle n \langle +\infty$, the chain–lattice system can also become ordered without the continuous phase transition [2]. We cannot say that all these phenomena are physical abnormalities for all possible values of $n$. As a matter of fact, they just show a common regular pattern, namely, agreement with their geometric topology and differential topology parameters

Last, the spin-projection orientations also depend on the dimensionality of the embeding space for the lattice system. If the embedded space is 2-dimensional, the spin-projection orientation is forbidden in the plane's normal direction for a plane lattice system or the circle-lattice system. The torus must be embedded in 3-dimensional space, and its non-vanishing vector field is also 2-dimensional at least. In the 1-dimensional space the spin-projection orientation normal to the real line cannot exist.

## 3.Conclusion

In summary, we found that the differential topology of the lattice-system of the Ising model determines whether there can be the continuous phase transition. The geometric topology of the space the lattice-system embedded in determines whether the system can become ordered. Also if the system behaves ordered it may not have a continuous phase transition. Thus the spin-projection orientations are strongly influenced by the geometric topology of the lattice-system embedding space in.